\documentclass[twocolumn,showpacs,preprintnumbers,amsmath,amssymb]{revtex4} 

\begin{document}
                                                                                
\preprint{CUPhys/10/2008}
                                                                                
\title{Phase Transition in a Long Range Antiferromagnetic Model}

\author{Anindita Ganguli }
\author{Subinay Dasgupta}%
\affiliation{%
Department of Physics, University of Calcutta,92 Acharya Prafulla Chandra 
Road, Calcutta 700009, India.\\
}%
                                                                                
\date{\today}

\begin{abstract}
We consider an Ising model where longitudinal components of every pair of spins
have antiferromagnetic interaction of the same magnitude. When subjected to
a transverse magnetic field at zero temperature, the system undergoes a phase
transition of second order to an ordered phase and if the 
temperature is now increased, there is another phase transition to disordered
phase. We provide derivation of these features by perturbative treatment up to the second order and argue that the results are non-trivial and not derivable from the known results about related models.

\end{abstract}


\def\be{\begin{equation}}
\def\ee{\end{equation}}
\maketitle

\section{Introduction} 

Ising model with long range 
interaction is of interest in statistical physics mainly in connection with the entanglement properties of the model and the quantum phase transitions displayed by it. A widely studied model of this type is the Lipkin-Meshkov-Glick (LMG) model (\cite{LMG}, \cite{RVM} and references therein) which deals with a system of spin-$\frac{1}{2}$ particles, each interacting with all other by the same strength and subjected to an external field. The general form of LMG Hamiltonian is
\be {\mathcal H}_{LMG} = - \frac{1}{N} \left[ J_x \left(S^x\right)^2 + J_y \left(S^y\right)^2 \right] - \Gamma S^z \ee
where $S^{\alpha}$ (for $\alpha =$ $x$, $y$, $z$) are the components  of total
magnetic moment,
\be S^{\alpha} = \sum_{j=1}^N s^{\alpha}_j,  \ee
$N$ is the number of spins, $J_x$, $J_y$ are the strength of 
interaction along X and Y directions for every pair 
of spins  and $\Gamma$ is the strength of an external field 
in the Z direction ($s^z = \pm 1/2$). This model has been attacked in various ways (\cite{RVM} and references therein). One way is to ignore the non-commutation of the operators $S^{\alpha}$ and treat them semi-classically :
\be S^x = {\mathcal S} \sin\theta\cos\phi, \;\;\; S^y = {\mathcal S} \sin\theta\sin\phi \;\;\; S^z = {\mathcal S} \cos\theta \ee
where ${\mathcal S}$ is the total spin. Another way \cite{OSDR} is to use Gaudin Lie algebra and 
derive an exact solution in the form of Bethe-like equations. 

In this paper, we consider a variant of LMG model which has been introduced recently and studied using quantum Monte Carlo method by Chandra, Inoue and Chakrabarti \cite{CIC}. One has an antiferromagnetic interaction of constant strength only in the longitudinal direction, and an external transverse field. The Hamiltonian is 
\be {\mathcal H} = \frac{J}{N} \left(S^z\right)^2 - \Gamma S^x \ee
with $J>0$. The LMG Hamiltonian reduces to this when either of $J_x$ and $J_y$ is zero and the other is negative. This Hamiltonian shows interesting phase transition properties as a function of field strength $\Gamma$ and temperature. Thus, when $\Gamma$ is zero, the system is in a highly
degenerate state with zero longitudinal (total) magnetic moment at zero as well
as non-zero temperature. But at zero temperature, as soon as a small 
$\Gamma$ is switched on, the system gets oriented completely in the transverse
direction. When temperature is increased, keeping $\Gamma$ non-zero, the
system undergoes a transition from an ordered to disordered state at a certain
critical temperature. The objective of this paper is to study the phase transition properties of the Hamiltonian $ {\mathcal H}$ using perturbative treatment. 
The first order term is zero and the second order term is calculated exactly.
In the next section we shall derive the
eigenvalues of ${\mathcal H}$ for {\em small} values of $\Gamma$ and in the
last section present some features of the result.

Before we conclude this section, we point out that the solution for the Hamiltonian 
${\mathcal H}$ does not follow from the known treatments of LMG Hamiltonian and hence the results of this paper are non-trivial. Since the commutator of the two terms in ${\mathcal H}$,
\be \left[ S_z^2 , S_x \right] = 2 i (S_y S_z + S_z S_y) \ee
has expectation value $\sim N$ when the spins are oriented in the transverse direction, we cannot make the semiclassical approximation described in Eq. (3). On the other hand, the parameter region covered here, although excluded by some discussions on LMG model \cite{RVM}, are indeed included in the exact solution of Ortiz et. al. \cite{OSDR}. However, the solution provided there is not in the form of a closed expression. We believe therefore, that the solution presented here are non-trivial. 

\section{Perturbative treatment of ${\mathcal H}$ } 

We shall study the Hamiltonian ${\mathcal H}$ for small values of 
$\Gamma$, by treating 
\be {\mathcal H}_p = - \Gamma S^x \ee
as the perturbation over the (unperturbed) Hamiltonian 
\be {\mathcal H}_0 = \frac{J}{N} \left(S^z\right)^2 \ee
The eigenvalues of ${\mathcal H}_0$ are obviously 
\be E^{(0)}_m = \frac{Jm^2}{N} \ee
with degeneracy 
\be \Omega_m = \frac{N !}
{\left(\frac{N}{2} + m\right)!\left(\frac{N}{2} - m\right)! } \ee
where,
\be m = - \frac{N}{2}, - \frac{N}{2}+1, \cdots \frac{N}{2}-1,\frac{N}{2}. \ee
Let us call the set of eigenstates with eigenvalue $E^{(0)}_m$ 
as ${\mathcal M}$. Since 
${\mathcal H}_p$ operating on any spin distribution $\mid~\!\!\!\alpha\rangle \in 
{\mathcal M}$ gives a state with a different $S^z$, the quantities
$\langle \beta \mid~{\mathcal H}_p \mid \alpha \rangle$ is zero for all
$\mid \alpha \rangle$, $\mid \beta \rangle$ $\in {\mathcal M}$. The first
order perturbation correction $E^{(1)}_m$ is hence nil and the dominant correction comes from the second order correction $E^{(2)}_m$. 

The second order perturbation correction $E^{(2)}_m$ to the eigenvalues
$E^{(0)}_m$ are the eigenvalues of a matrix ${\bf P}$ whose elements are given by
\be P_{\alpha \beta} = \sum_l \frac
{\langle \alpha \mid {\mathcal H}_p \mid l \rangle 
\langle l \mid {\mathcal H}_p \mid \beta \rangle }{E^{(0)}_m - E^{(0)}_l} \ee
where $\mid \alpha \rangle$, $\mid \beta \rangle$ $\in {\mathcal M}$ but 
$\mid l \rangle$ $\notin {\mathcal M}$. We shall now see that the eigenproblem of ${\bf P}$ cn be solved exactly. First consider the diagonal element
$P_{\alpha \alpha}$. The state $\mid \alpha \rangle$ has $(N/2)+m$ up-spins and
$(N/2)-m$ down-spins. When the operator ${\mathcal H}_p$ flips an up/down spin
$E^{(0)}_l$ becomes $(m\mp 1)J$, so that
\be P_{\alpha \alpha} = \frac{N \Gamma^2}{4J} \frac{4m^2 + N}{4m^2 - 1} \ee
As regards the off-diagonal elements $P_{\alpha \beta}$, we note that it is 
non-zero when and only when the spin-distributions $\mid \alpha \rangle$
and $\mid \beta \rangle$ differ in precisely {\em two unlike} spins (since
these two states must have the same $S^z$). Thus, the two spin-states will be 
like :
\[ \mid \alpha \rangle\; : \; \mid \cdots + \cdots - \cdots  \rangle \;\;\;\;\;
   \mid \beta  \rangle\; : \; \mid \cdots - \cdots + \cdots  \rangle \]
Then $\mid l \rangle$ can be either $\mid \cdots + \cdots + \cdots  \rangle$
(with $E^{(0)}_l = (m+1)J$) or $\mid \cdots - \cdots - \cdots  \rangle$
(with $E^{(0)}_l = (m-1)J$). Hence, $P_{\alpha \beta}$, if non-zero will have
the value 
\be P_{\alpha \beta} = \frac{N \Gamma^2}{2J} \frac{1}{4m^2 - 1} \ee

To proceed, let us now consider the Hamiltonian
\be {\mathcal H}^{\prime} = \frac{J}{N} \left(S^z\right)^2 
+ \frac{h}{N} \left[ \left(S^x\right)^2 + \left(S^y\right)^2 \right] \ee
and treat 
\be {\mathcal H}_p^{\prime} = \frac{h}{N} \left[ 
\left(S^x\right)^2 + \left(S^y\right)^2 \right] \ee
as perturbation on the (unperturbed) Hamiltonian $ \frac{J}{N} \left(S^z\right)^2$ which is nothing but the ${\mathcal H}_0$ of Eq. (5).
The first order perturbation matrix ${\bf P^{\prime}}$ has diagonal elements
\begin{eqnarray}
P^{\prime}_{\alpha \alpha} & \equiv & 
\langle \alpha \mid {\mathcal H}_p^{\prime} \mid \alpha \rangle \nonumber \\
& = & \langle \alpha \mid \frac{h}{N} \sum_{j,k}
\left(s_j^x s_k^x + s_j^y s_k^y\right) \delta_{j,k}\mid \alpha \rangle 
\nonumber \\
& =& h/2 \end{eqnarray}
The off-diagonal elements $P^{\prime}_{\alpha \beta}$ will be $h/N$ if 
$\mid \alpha \rangle$
and $\mid \beta \rangle$ differ in precisely two unlike spins and will be zero
otherwise. Thus, we get the correspondence
\be {\bf P} =  \frac{N \Gamma^2}{4J} \frac{4m^2 + N}{4m^2 - 1} {\bf 1} + 
\frac{N^2 \Gamma^2}{2Jh} \frac{1}{4m^2 - 1}\left({\bf P^{\prime}}
- \frac{h}{2}{\bf 1}\right) \ee
where ${\bf 1}$ is the $\Omega_m \times \Omega_m$ unit matrix.

It is trivial to diagonalise the Hamiltonian ${\mathcal H}^{\prime}$ by
rewriting it as,
\be {\mathcal H}^{\prime} = \frac{J-h}{N} \left(S^z\right)^2 
+ \frac{h}{N} \left(\hat{{\mathcal S}}\right)^2 \ee
(where $\hat{{\mathcal S}}$ is the total spin operator) and noting that the
eigenstates of this Hamiltonian is labelled by $m$ and the total spin ${\mathcal S}$. Using the usual relationships
\[ \hat{{\mathcal S}}^2 \mid {\mathcal S},m \rangle = {\mathcal S}({\mathcal S}
+1) \mid {\mathcal S},m \rangle \]
\[ {\mathcal S} = 0, 1, 2, \cdots \frac{N}{2} \]
and 
\[ m = - {\mathcal S}, - {\mathcal S}+1, \cdots {\mathcal S}-1, {\mathcal S} \]
we observe that the eigenvalues of ${\bf P^{\prime}}$ are
\be \frac{h}{N}[{\mathcal S}({\mathcal S}+1) - m^2] \ee

From Eq.(17) we can finally conclude that the unperturbed eigenstate of 
${\mathcal H}_0$ with eigenvalue $(J/N)m^2$ splits up under second-order
perturbation by ${\mathcal H}_p$ to eigenstates with eigenvalues
\be \lambda({\mathcal S}, m) =  \frac{N \Gamma^2}{4J(4m^2 - 1)}\left[
2m^2 + 2{\mathcal S}({\mathcal S}+1) \right] \ee
The degeneracy of this state can be easily seen to be \cite{dicke,vidal}
\begin{eqnarray} 
D({\mathcal S}, m) & = & \frac{N !}
{\left(\frac{N}{2} + {\mathcal S}\right)!\left(\frac{N}{2} -
{\mathcal S}\right)! } - \nonumber \\
 & & \frac{N !}
{\left(\frac{N}{2} + {\mathcal S} + 1\right)!\left(\frac{N}{2} -
{\mathcal S} - 1\right)! } \end{eqnarray}

\section{Discussion}

Having solved the eigenproblem of ${\mathcal H}$ upto second order in $\Gamma$, we can now observe the 
following features : (i) At zero transverse field, the system has Hamiltonian
${\mathcal H}_0$ and is in a state of $m=0$ for zero or non-zero temperature.
(ii) At zero temperature, when a small (transverse) field is turned on, 
the ground state energy (correct upto $\Gamma^2$) becomes,
\be E^{(2)}_0 = - \frac{N \Gamma^2}{2J} {\mathcal S}({\mathcal S}+1) \ee
with ${\mathcal S} = N/2$. The susceptibility per spin,
\be \chi \equiv \frac{1}{N}\left(\frac{d^2E^{(2)}_0}{d\Gamma^2}\right)_
{\Gamma=0} = - \frac{N^2}{8J}\left(1 + \frac{2}{N} \right) \ee
diverges in the thermodynamic limit, indicating a second order phase transition
at $\Gamma=0$. (iii) {\em At non-zero temperature} $T$, the free energy 
(correct upto $\Gamma^2$) is,
\be F^{(2)}_0 = - \frac{N \Gamma^2}{2J} {\mathcal S}({\mathcal S}+1)
- k_B T \ln D({\mathcal S}, m) \ee
where $k_B$ is Boltzmann constant. The value of ${\mathcal S}$ which minimises
this expression is hence the solution of the equation
\be x = \tanh(\alpha x) \ee
where $x = 2{\mathcal S}/N$, and $\alpha = N^2\Gamma^2/(4Jk_BT)$. To arrive at 
this relation we have assumed $N$, ${\mathcal S}$ and $(N/2 - {\mathcal S})$
to be large and used Stirling's approximation for the factorials of Eq.(21). This shows that
there is a mean-field type phase transition at a critical temperature
\be \frac{k_B T_c}{J} =  \frac{1}{4}\left(\frac{N\Gamma}{J}\right)^2. \ee
Two aspects of this phase transition are to be noted. (i) Since 
the expectation value of $\left(S^x\right)^2$ in the unperturbed state is,
\[ \langle {\mathcal S}, m \mid \left(S^x\right)^2 \mid {\mathcal S}, m \rangle
 = \frac{1}{2} ({\mathcal S}^2 + {\mathcal S} - m^2), \]
at $T > T_c$ (or $\alpha < 1$), there is disordered phase with
${\mathcal S}=0$ and $\langle \left(S^x\right)^2 \rangle =0$, 
while for $T < T_c$, we have an ordered phase with ${\mathcal S} \ne 0$ and 
$\langle \left(S^x\right)^2 \rangle \ne0$. (ii) The appearance of $N$ in the expression of $T_c$ is uncomfortable at the first sight.
Note that for the perturbative treatment to be valid,
the perturbation correction must be less than the spacing between two adjacent
eigenvalues of the unperturbed Hamiltonian. According to Eq. (4) this needs,
\be \Gamma \ll J/N \ee 
so that 
\be \frac{k_B T_c}{J} \ll 1. \ee
Thus, all we can claim is that when the condition (27) is fulfilled there will be a temperature-induced phase transition at a temperature given by Eq. (26). In other words, if we rewrite the Hamiltonian of Eq. (4) as  
\be {\mathcal H}_s = \left(S^z\right)^2 - b S^x  \ee
(where $b = \Gamma N/J$), then for $b=0$, there is no critical temperature but when $b$ has a  {\em small} non-zero value, there is a critical temperature given by $T_c = b^2/4$ and this conclusion is true for all values of $N$.
Our treatment cannot predict anything when $b$ is {\em not} small. In the finite size study of 
Chandra, Inoue and Chakrabarti \cite{CIC} the value of $b$ is 100 and hence the fact that this study does not observe a thermal phase transition does not present a contradiction with our results. 

We are grateful to I. Bose, J. Inoue, G. Ortiz and D. Sen for helpful discussions and encouragement. 
One author (AG) is grateful to UGC for UPE scholarship. The work was financed
by UPE Grant (Computational Group) and by CSIR project.

\end{document}